\documentclass[11pt]{article}
\usepackage[utf8]{inputenc}
\usepackage[T1]{fontenc}
\usepackage{mathptmx}
\usepackage{amsmath,amssymb}
\usepackage[margin=1in]{geometry}
\usepackage{graphicx}
\setkeys{Gin}{keepaspectratio}
\usepackage{longtable,booktabs,array,calc}
\usepackage[labelsep=period,font=small]{caption}
\usepackage{authblk}
\usepackage[hidelinks]{hyperref}
\usepackage{microtype}
\providecommand{\tightlist}{\setlength{\itemsep}{0pt}\setlength{\parskip}{0pt}}

\title{Field-aligned measurement removes the anisotropy barrier of sample-based quantum subspace solvers for magnetized-plasma heat conduction}
\author[1,2]{Doyeol (David) Ahn\thanks{Author to whom correspondence should be addressed: dahn@uos.ac.kr}}
\affil[1]{Department of Electrical and Computer Engineering, University of Seoul, 163 Seoulsiripdae-ro, Dongdaemun-gu, Seoul 02504, Republic of Korea}
\affil[2]{Singularity Quantum Inc., Newport Beach, California, USA}
\date{}

\begin{document}
\maketitle
\begin{abstract}
Sample-based quantum subspace methods construct a small subspace from measured basis states and solve the target problem classically within it. We study these methods for strongly anisotropic heat conduction in magnetized plasmas, where the ratio of parallel to perpendicular conductivity reaches \(10^{8}\)--\(10^{10}\). We prove two results for the energy-norm Galerkin solution. First, if the measurement basis is aligned with the flux surfaces and the subspace contains the poloidal-mode-zero block, the error is bounded by \(\|r_0\|_{L^+}(\chi_\parallel-\chi_\perp)^{-1/2}\), where the numerator does not depend on \(\chi_\parallel\). This bound is asymptotically sharp, so the accuracy improves as the anisotropy grows. Second, in the computational (nodal) basis, a subspace that misses a single node of a flux surface suppresses the flux-function component on that surface as \(\chi_\parallel^{-1/2}\). If the limiting solution is nonzero on such a surface, the error has a floor that does not vanish as \(\chi_\parallel\to\infty\). We verify both results on a shaped cross-section and on a KSTAR-like Miller equilibrium with ten qubits. At \(\chi_\parallel/\chi_\perp=10^9\), the nodal basis gives an energy-norm error of order unity for every shot budget, whereas a quantum Fourier transform on the poloidal register gives \(1.3\times10^{-5}\) with \(2.5\times10^{4}\) noiseless samples. The same barrier limits sample-based quantum diagonalization with configuration-recovery-type support expansion, and a residual-norm formulation fails in double precision because it squares the condition number. We also show that the aligned measurement requires no time evolution, which reduces the circuit to state preparation and a five-qubit quantum Fourier transform. On a superconducting processor with ten qubits, the aligned basis gives a median error of \(7.0\times10^{-6}\) at \(\chi_\parallel/\chi_\perp=10^9\) for subsamples of 768 shots, whereas the computational basis gives an error above 0.94 up to \(1.2\times10^4\) shots. The same logical circuits on a trapped-ion processor give the same behavior.
\end{abstract}

\hypertarget{i.-introduction}{%
\section*{I. Introduction}}

Heat conduction in a magnetized plasma is strongly anisotropic {[}1,2{]}. The parallel conductivity \(\chi_\parallel\) exceeds the perpendicular conductivity \(\chi_\perp\) by eight to ten orders of magnitude in the core of a fusion device {[}1{]}. This anisotropy determines the temperature profile, and it enters the computation of equilibria, stability, and fusion power. On the numerical side, the problem is stiff. The condition number of the discrete operator grows in proportion to \(\chi_\parallel/\chi_\perp\), and discretizations that are not aligned with the magnetic field produce spurious perpendicular diffusion {[}3{]}. Asymptotic-preserving methods address this difficulty by separating the field-line-averaged part of the solution {[}4{]}.

Quantum algorithms for partial differential equations have been proposed for plasma problems {[}5{]} and for nonlinear fluid equations {[}6{]}. Among them, sample-based quantum subspace methods are attractive for present hardware. Sample-based quantum diagonalization (SQD) samples configurations on a quantum processor, repairs the samples, and diagonalizes the Hamiltonian projected onto the span of the configurations {[}7{]}. Sample-based Krylov quantum diagonalization (SKQD) uses quantum Krylov states as the sampling states {[}8{]}. The same strategy has been extended to linear systems, either through a cost Hamiltonian {[}9{]} or through an energy-norm Galerkin formulation {[}10{]}, and code-space recovery has been introduced for problems without a native symmetry constraint {[}11{]}. Iterative refinement has also been used to obtain converged solutions of a Poisson problem on superconducting and trapped-ion hardware {[}12{]}.

These methods share one assumption. The target state must be compressible in the measurement basis, and the sampling state must assign appreciable probability to the relevant support {[}11{]}. Here, we ask whether this assumption holds for anisotropic heat conduction, and what can be done when it fails. The answer depends on the measurement basis rather than on the number of shots.

The key object is the poloidal-mode-zero block. In flux coordinates \((\rho,\theta)\), a temperature that is constant on every flux surface has only the poloidal Fourier mode \(m=0\). The parallel conduction operator annihilates exactly these functions, so they span the near-null space of the stiff part of the operator. We call the subspace spanned by the \(m=0\) mode on each flux surface the \(m=0\) block. As \(\chi_\parallel/\chi_\perp\) grows, the solution approaches a function in this block, and the remaining poloidal variation decreases as \(\chi_\perp/\chi_\parallel\). A measurement basis that resolves the poloidal modes therefore exposes the \(m=0\) block directly, whereas the computational basis spreads it over all nodes of a surface.

In order to test the results in a realistic tokamak geometry, we use the Miller equilibrium model {[}13{]}. Miller \emph{et al.} parametrized the flux surfaces of a shaped tokamak by the elongation \(\kappa(\rho)\), the triangularity \(\delta(\rho)\), and the Shafranov shift \(\Delta(\rho)\) of each surface, together with the major and minor radii. This local equilibrium model reproduces the shape of measured equilibria with a few parameters, and it is widely used in gyrokinetic and transport calculations. We choose the parameters close to those of KSTAR.

The contributions of this paper are as follows.

\begin{enumerate}
\def\labelenumi{(\arabic{enumi})}
\item
  We prove a sharp upper bound for the Galerkin error in a field-aligned measurement basis. The error decreases as \((\chi_\perp/\chi_\parallel)^{1/2}\), and the constant does not depend on \(\chi_\parallel\) (Theorem 1).
\item
  We prove that the computational basis has an anisotropy barrier. A subspace that misses one node of a flux surface cannot represent the flux-function component on that surface as \(\chi_\parallel\to\infty\), and the error has a floor that does not vanish (Theorem 2).
\item
  We verify both results numerically on a shaped cross-section and on a KSTAR-like Miller equilibrium with ten qubits, for \(\chi_\parallel/\chi_\perp\) up to \(10^9\).
\item
  We show that the barrier also limits SQD with configuration-recovery-type support expansion, and that the residual-norm formulation of the linear solver fails in double precision because it squares the condition number. The energy-norm formulation does not have this defect.
\item
  We show that the aligned measurement requires no time evolution. The resulting circuit consists of state preparation and a quantum Fourier transform on the poloidal register.
\item
  We demonstrate the effect on a superconducting quantum processor (ibm\_cleveland) with ten qubits. With subsamples of 768 shots, the \(\theta\)-QFT basis gives a median energy-norm error of \(7.0\times10^{-6}\) at \(\chi_\parallel/\chi_\perp=10^9\), and the error is consistent with the \((\chi_\perp/\chi_\parallel)^{1/2}\) scaling of Theorem 1 over the tested range. The computational basis gives an error above 0.94 up to \(1.2\times10^4\) shots on the same device. The same logical circuits on a trapped-ion processor (IonQ Forte-1) give \(9.9\times10^{-6}\).
\end{enumerate}

The paper is organized as follows. Section II describes the model and its discretization. Section III describes the sample-based subspace solver. Section IV gives the two theorems. Section V presents the numerical results, and Sec. VI compares the other sample-based methods. Section VII describes the hardware realization. Section VIII states the limitations, and Sec. IX concludes. Appendix A lists the hardware jobs.

\hypertarget{ii.-model-and-discretization}{%
\section*{II. Model and discretization}}

\hypertarget{a.-anisotropic-heat-conduction}{%
\subsection*{A. Anisotropic heat conduction}}

We consider the steady equation

\[-\nabla\cdot\left(\mathsf{D}\,\nabla T\right) = Q, \qquad \mathsf{D} = \chi_\parallel\,\mathbf{b}\mathbf{b} + \chi_\perp\left(\mathsf{I}-\mathbf{b}\mathbf{b}\right), \tag{1}\]

where \(T\) is the temperature, \(\mathbf{b}\) is the unit vector along the magnetic field, and \(Q\) is the heat source. We impose \(T=0\) on the boundary. In a tokamak, the field consists of a toroidal part \(B_\phi\mathbf{e}_\phi\) and a poloidal part \(\mathbf{B}_p\), which lies in the poloidal \((R,Z)\) plane. We write \(\mathbf{b}=(B_\phi\mathbf{e}_\phi+B_p\mathbf{b}_p)/B\), where \(\mathbf{b}_p=\mathbf{B}_p/B_p\) is the unit vector of the poloidal projection. For an axisymmetric temperature, \(\partial T/\partial\phi=0\), and \(\mathbf{b}\cdot\nabla T=(B_p/B)\,\mathbf{b}_p\cdot\nabla T\). The toroidal component of the heat flux does not contribute to the divergence. By substituting this into (1), we obtain the two-dimensional equation in the poloidal plane with
\[\mathsf{D}_p = \chi_\perp\mathsf{I} + \left(\chi_\parallel-\chi_\perp\right)\left(\frac{B_p}{B}\right)^2\mathbf{b}_p\mathbf{b}_p .\]
Therefore, only the poloidal projection \(\mathbf{b}_p\) enters, and the parallel conductivity is reduced to \(\chi_{\parallel,\mathrm{eff}} = \chi_\perp+(\chi_\parallel-\chi_\perp)(B_p/B)^2\). We take \(B_p/B\) as constant. In the following, \(\chi_\parallel\) denotes this effective value, and \(\mathbf{b}\) denotes \(\mathbf{b}_p\).

\hypertarget{b.-non-aligned-discretization}{%
\subsection*{B. Non-aligned discretization}}

In order to show the difficulty of non-aligned grids, we first use the benchmark of Günter \emph{et al.} {[}3{]}. The domain is \([-1/2,1/2]^2\), the field lines are the contours of \(\psi=\cos\pi x\cos\pi y\), and \(Q = 2\pi^2\chi_\perp\psi\). The exact solution is \(T=\psi\) for every \(\chi_\parallel\). We discretize (1) with continuous tensor-product Lagrange elements of degree \(p=1\)--\(4\) (\(Q_p\) elements). The domain is divided into \(N_e\times N_e\) square elements. On each element, the basis functions are products of one-dimensional Lagrange polynomials of degree \(p\) in \(x\) and in \(y\). Their nodes are the \(p+1\) Gauss--Lobatto--Legendre points in each direction, and the basis functions are continuous across element boundaries. The case \(p=1\) is the bilinear element, and \(p=2\) is the biquadratic element. A higher degree resolves the curved field lines inside each element more accurately at the same number of unknowns. The number of interior unknowns is \(n=(N_ep-1)^2\), and the integrals are evaluated by Gauss quadrature. The numerical perpendicular diffusion is measured by \(\chi_{\perp,\mathrm{num}}/\chi_\perp = 1/T(0,0)-1\).

\hypertarget{c.-field-aligned-discretization}{%
\subsection*{C. Field-aligned discretization}}

We introduce flux coordinates \((\rho,\theta)\), where \(\rho\) labels the flux surface and \(\theta\) is the poloidal angle. The cross-section is described by the map \((\rho,\theta)\mapsto(R,Z)\). We use two maps. The first is a shaped section,

\[R = \rho\cos(\theta+\delta\sin\theta), \qquad Z = E\rho\sin\theta, \tag{2}\]

with \(E=1.5\) and \(\delta=0.3\). The second is a Miller-type equilibrium {[}13{]},

\[R = R_0 + \Delta(\rho) + a\rho\cos\left[\theta + \arcsin\delta(\rho)\,\sin\theta\right], \qquad Z = \kappa(\rho)\, a\rho\sin\theta, \tag{3}\]

with \(R_0 = 1.8\) m, \(a = 0.5\) m, \(\kappa(\rho) = 1+0.8\rho^2\), \(\delta(\rho) = 0.5\rho^2\), and \(\Delta(\rho) = 0.08a(1-\rho^2)\). These values are close to the KSTAR geometry. For (3) the volume element contains the toroidal factor \(2\pi R\), and \(\chi_\perp(\rho) = 1+4\rho^2\). In both cases \(\rho\in[0.2,1]\). Figure 1 shows the coordinate lines of both maps and the profiles of the Miller parameters. The shaped section (2) is a planar model, whereas (3) is a toroidal section with \(R\) and \(Z\) in meters.

\begin{figure}
\centering
\includegraphics[width=1\textwidth,height=\textheight]{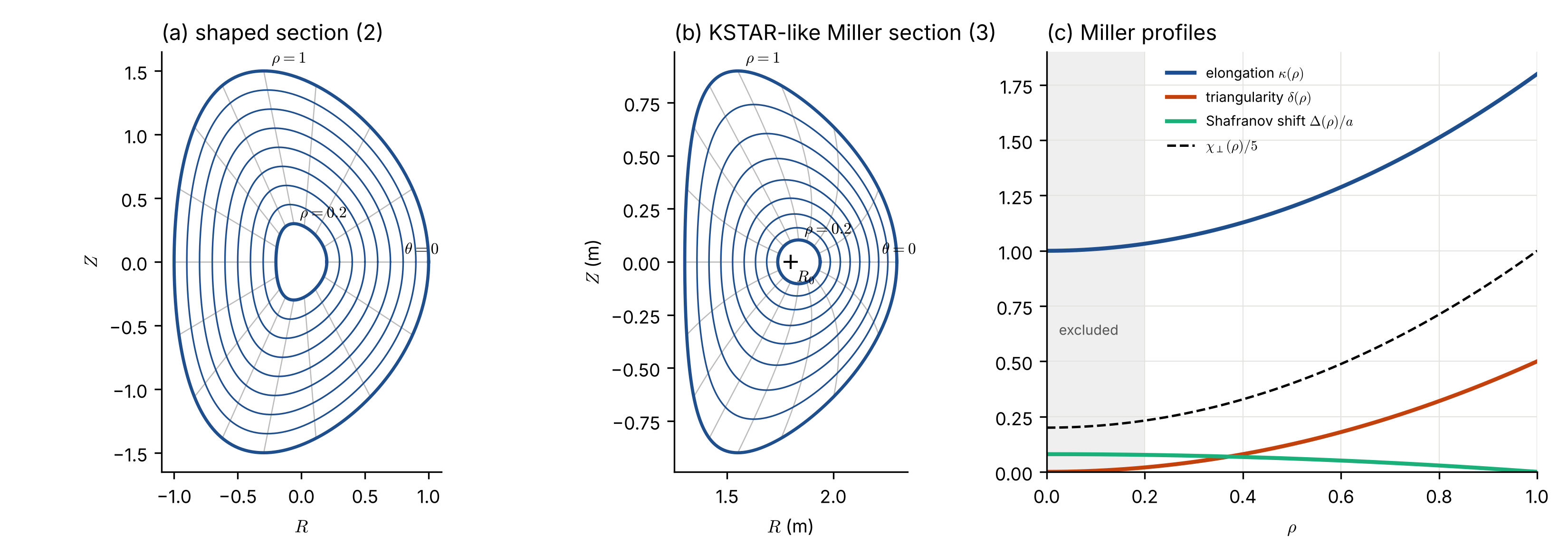}
\caption{Flux coordinates. (a) Shaped section (2) with \(E=1.5\) and \(\delta=0.3\). (b) KSTAR-like Miller section (3) with \(R_0=1.8\) m and \(a=0.5\) m; the cross marks the geometric axis \(R_0\). Blue curves: flux surfaces \(\rho=0.2,0.3,\dots,1\); gray lines: \(\theta=\mathrm{const}\). The region \(\rho<0.2\) is excluded. (c) Elongation, triangularity, Shafranov shift, and perpendicular conductivity of the Miller section.}
\end{figure}

The field is tangent to the surfaces \(\rho=\mathrm{const}\), so that \(\mathbf{b}\cdot\nabla T = T_\theta/\sqrt{g_{\theta\theta}}\). Here \(g_{\theta\theta}\) denotes the covariant metric component. The covariant metric tensor of the map \((\rho,\theta)\mapsto(R,Z)\) and its inverse, the contravariant metric tensor, are
\[
g_{ij}=\begin{pmatrix} R_\rho^2+Z_\rho^2 & R_\rho R_\theta+Z_\rho Z_\theta\\ R_\rho R_\theta+Z_\rho Z_\theta & R_\theta^2+Z_\theta^2\end{pmatrix},\qquad
g^{ij}=\left(g_{ij}\right)^{-1}=\frac{1}{J_p^2}\begin{pmatrix} g_{\theta\theta} & -g_{\rho\theta}\\ -g_{\rho\theta} & g_{\rho\rho}\end{pmatrix},
\]
where the subscripts \(\rho\) and \(\theta\) on \(R\) and \(Z\) denote partial derivatives, and \(J_p=R_\rho Z_\theta-R_\theta Z_\rho=(\det g_{ij})^{1/2}\) is the Jacobian in the poloidal plane. The contravariant components are equal to \(g^{\rho\rho}=|\nabla\rho|^2\), \(g^{\rho\theta}=\nabla\rho\cdot\nabla\theta\), and \(g^{\theta\theta}=|\nabla\theta|^2\). The volume element is \(J\,d\rho\,d\theta\) with \(J=J_p\) for (2) and \(J=2\pi RJ_p\) for (3). The energy of a temperature field is

\[a(T,T) = \int\left[\chi_\perp\left(g^{\rho\rho}T_\rho^2 + 2g^{\rho\theta}T_\rho T_\theta + g^{\theta\theta}T_\theta^2\right) + \left(\chi_\parallel-\chi_\perp\right)\frac{T_\theta^2}{g_{\theta\theta}}\right] J\, d\rho\, d\theta, \tag{4}\]

where \(g^{ij}\) are the contravariant metric components and \(J\) is the Jacobian. We use \(N_\rho\) uniform interior nodes in \(\rho\) and \(N_\theta\) uniform nodes in \(\theta\). The gradients are evaluated at cell centers \((\rho_{i+1/2},\theta_{j+1/2})\) by forward differences and averages, and (4) is integrated by the midpoint rule. As a result, we obtain a symmetric positive-definite system

\[A x = f, \qquad A = A_\perp + (\chi_\parallel-\chi_\perp)\,L. \tag{5}\]

Here \(A_\perp\) is the perpendicular operator with conductivity \(\chi_\perp\) in all directions, and \(L\succeq 0\) is the parallel operator per unit conductivity.

The forward difference in \(\theta\) annihilates only the grid functions that do not depend on \(\theta\). It can be shown that the kernel of \(L\) is

\[\mathcal{N} = \ker L = \{x : x_{ij} = u_i \ \text{for all}\ j\}, \qquad \dim\mathcal{N} = N_\rho. \tag{6}\]

In the \(\theta\)-Fourier basis, \(\mathcal{N}\) is exactly the \(m=0\) block. We verified numerically that \(L\) has exactly \(N_\rho = 32\) zero eigenvalues and that the coupling between the \(m=0\) block and the other modes is of the order of \(10^{-16}\), the rounding level. For the source (7), the \(\theta\) variation of the discrete solution is \(2.5\times10^{-2}\), \(3.2\times10^{-4}\), \(3.2\times10^{-6}\), and \(3.3\times10^{-9}\) at \(\chi_\parallel/\chi_\perp = 10^2\), \(10^4\), \(10^6\), and \(10^9\). It is therefore proportional to \(1/\chi_\parallel\), and no spurious perpendicular diffusion appears.

We use \(N_\rho=N_\theta=32\) (\(n=1024\), ten qubits) unless noted otherwise. The index of an unknown is \(i N_\theta + j\), so that the \(\rho\) register holds the five most significant qubits and the \(\theta\) register holds the five least significant qubits. The source consists of a background and an off-axis localized deposition,

\[Q = Q_0(\rho) + Q_1 \exp\left[-\frac{(R-R_1)^2+(Z-Z_1)^2}{2w^2}\right], \tag{7}\]

which breaks the poloidal symmetry. For (2) we use \(Q_0=1\), \(Q_1=50\), \((R_1,Z_1)=(0.45,0.35)\), and \(w=0.08\). For (3) we use \(Q_0 = 2(1-\rho^2)\), \(Q_1=40\), \((R_1,Z_1)=(R_0+0.30,\,0.30)\), and \(w=0.06\), which represents an electron-cyclotron heating spot.

\hypertarget{iii.-sample-based-subspace-solver}{%
\section*{III. Sample-based subspace solver}}

A sample-based subspace solver proceeds in three steps. First, a sampling state is prepared on the quantum processor. In the general form of the method, the sampling states are the Krylov states \(|\phi_k\rangle = e^{-ik\Delta t\,\hat A}|f\rangle\) with \(\hat A = A/\|A\|\), \(k=0,\dots,K-1\), and \(\Delta t = 1\). The case \(K=1\) is source-state sampling, in which only \(|f\rangle\) is prepared. We use Krylov sampling in Sec. V and show in Sec. V E that source-state sampling is sufficient in the aligned basis. The hardware experiments of Sec. VII use source-state sampling. Second, each state is measured in a chosen basis \(\{|s\rangle\}\), and the set \(S\) of observed outcomes is collected. Third, the Galerkin solution is computed classically on \(\mathcal{V}_S = \mathrm{span}\{|s\rangle : s\in S\}\),

\[x_S = V_S\left(V_S^T A V_S\right)^{-1}V_S^T f, \tag{8}\]

where the columns of \(V_S\) are the basis vectors in \(S\). The Galerkin solution is the best approximation of \(x^*\) in \(\mathcal{V}_S\) in the energy norm \(\|e\|_A = (e^TAe)^{1/2}\). We report the relative error \(\|x^*-x_S\|_A/\|x^*\|_A\).

We compare two measurement bases. The computational basis measures the nodal values directly. The aligned basis applies a quantum Fourier transform (QFT) to the \(\theta\) register before measurement, followed by a fixed recombination of the \(\pm m\) pairs into real cosine and sine modes. The \(\rho\) register is measured directly.

\hypertarget{iv.-theorems}{%
\section*{IV. Theorems}}

Throughout this section, we assume that the grid is fixed, that \(A_\perp\) is symmetric positive definite, that \(L\) is symmetric positive semidefinite with \(\ker L=\mathcal{N}\), and that \(\alpha=\chi_\parallel-\chi_\perp>0\). Then \(A=A_\perp+\alpha L\), and \(A\succeq\alpha L\). We denote by \(\lambda_{\mathrm{gap}}\) the smallest nonzero eigenvalue of \(L\).

\hypertarget{a.-aligned-measurement-basis}{%
\subsection*{A. Aligned measurement basis}}

Let \(V_0\) be an orthonormal basis of \(\mathcal{N}\), and let \(u_0 = V_0(V_0^TAV_0)^{-1}V_0^Tf\) be the Galerkin solution on \(\mathcal{N}\).

\textbf{Theorem 1.} The vector \(u_0\) and the residual \(r_0 = f - A_\perp u_0\) do not depend on \(\chi_\parallel\). For every subspace \(\mathcal{V}_S\supseteq\mathcal{N}\),

\[\|x^*-x_S\|_A \le \frac{\|r_0\|_{L^+}}{\sqrt{\chi_\parallel-\chi_\perp}}, \qquad \|r_0\|_{L^+} = \left(r_0^TL^+r_0\right)^{1/2}, \tag{9}\]

where \(L^+\) is the pseudo-inverse of \(L\).

\emph{Proof.} Since \(Lu_0=0\), we have \(Au_0 = A_\perp u_0\). Therefore \(u_0\) solves \(V_0^TA_\perp V_0\,c = V_0^Tf\), which contains no \(\chi_\parallel\), and \(r_0 = f-Au_0\). The Galerkin condition gives \(V_0^Tr_0=0\), so that \(r_0\) lies in the range of \(L\) and \(r_0 = LL^+r_0\). Let \(\delta = x^*-u_0\). Then \(A\delta = r_0\), and

\[\delta^TA\delta = \delta^Tr_0 = \delta^TLL^+r_0 \le \left(\delta^TL\delta\right)^{1/2}\|r_0\|_{L^+}. \tag{10}\]

Since \(A\succeq(\chi_\parallel-\chi_\perp)L\), we obtain \(\delta^TL\delta\le\delta^TA\delta/(\chi_\parallel-\chi_\perp)\). By substituting this into (10), we obtain \(\|\delta\|_A\le\|r_0\|_{L^+}/\sqrt{\chi_\parallel-\chi_\perp}\). The Galerkin solution on \(\mathcal{V}_S\supseteq\mathcal{N}\) is not worse than \(u_0\) in the energy norm. \(\square\)

\textbf{Corollary 1.} If \(r_0\neq0\), the bound (9) is attained asymptotically for \(\mathcal{V}_S=\mathcal{N}\),
\[\frac{\|x^*-u_0\|_A^2}{\|r_0\|_{L^+}^2/\alpha} = 1-O(\alpha^{-1}), \qquad \alpha\to\infty. \tag{11}\]

\emph{Proof.} We write vectors in the orthogonal decomposition \(\mathcal{N}\oplus\mathcal{N}^\perp\), so that \(r_0=(0,r)\) and \(L=\mathrm{diag}(0,L_{22})\) with \(L_{22}\succ0\). Let \(C\) be the Schur complement of the \(\mathcal{N}\) block in \(A_\perp\), which is symmetric positive definite. Block elimination gives \(\|x^*-u_0\|_A^2=\delta^Tr_0=r^T(C+\alpha L_{22})^{-1}r\). Since \(L_{22}^{-1/2}CL_{22}^{-1/2}\succ0\), we obtain \(\alpha\,r^T(C+\alpha L_{22})^{-1}r = r^TL_{22}^{-1}r - O(\alpha^{-1})\), with \(r^TL_{22}^{-1}r=\|r_0\|_{L^+}^2\). \(\square\)

The corollary holds for every \(A_\perp\), \(L\), and \(f\) that satisfy the assumptions above. Since \(C\succ0\), we have \((C+\alpha L_{22})^{-1}\prec(\alpha L_{22})^{-1}\), so that the ratio in (11) is less than one for every \(\alpha\), which is consistent with (9).

We now consider the number of shots needed to collect \(\mathcal{N}\). In the aligned basis, \(\mathcal{N}\) consists of \(N_\rho\) basis states. If each of them is observed in every shot with a probability of at least \(p_{\min}>0\), a union bound shows that \(p_{\min}^{-1}\ln(N_\rho/\epsilon)\) shots collect all of them with a probability of at least \(1-\epsilon\). For source-state sampling, \(p_i\) is the probability of the \(m=0\) mode of \(f\) on surface \(i\). It does not depend on \(\chi_\parallel\), because \(f\) does not, and it is positive when \(f\) has a nonzero average on every surface. The number of shots is then independent of \(\chi_\parallel\). For Krylov sampling, \(p_i\) depends on \(\chi_\parallel\) through the evolution, and the same statement holds whenever \(p_{\min}\) is bounded away from zero.

\hypertarget{b.-computational-basis}{%
\subsection*{B. Computational basis}}

\textbf{Theorem 2.} Let \(S\) be a set of grid nodes and let \(\mathcal{I}\) be the set of flux surfaces that contain at least one node outside \(S\). Let \(P_\perp\) be the orthogonal projector onto \(\mathcal{N}^\perp\). For every \(v\in\mathrm{span}\{e_s:s\in S\}\) and every \(i\in\mathcal{I}\),

\[|\bar v_i| \le \|P_\perp v\| \le \left(\frac{v^TAv}{(\chi_\parallel-\chi_\perp)\lambda_{\mathrm{gap}}}\right)^{1/2}, \tag{12}\]

where \(\bar v_i\) is the mean of \(v\) on surface \(i\). Consequently, with \(\eta = \|x^*\|_A/\sqrt{(\chi_\parallel-\chi_\perp)\lambda_{\mathrm{gap}}}\),

\[\|x^*-x_S\|_A \ge \left(\lambda_{\min}(A_\perp)\,N_\theta\sum_{i\in\mathcal{I}}\left(|\bar x^*_i|-\eta\right)_+^2\right)^{1/2}. \tag{13}\]

\emph{Proof.} The orthogonal projection of \(v\) onto \(\mathcal{N}\) takes the value \(\bar v_i\) at every node of surface \(i\). Let \(j_0\) be a node of surface \(i\) outside \(S\). Then \(v_{j_0}=0\), and \(|\bar v_i| = |(P_\perp v)_{j_0}|\le\|P_\perp v\|\). The second inequality in (12) follows from \(v^TAv\ge(\chi_\parallel-\chi_\perp)v^TLv\ge(\chi_\parallel-\chi_\perp)\lambda_{\mathrm{gap}}\|P_\perp v\|^2\). For (13), we use \(x_S^TAx_S\le x^{*T}Ax^*\), which holds for the Galerkin solution. Then the surface means of \(x_S\) on \(\mathcal{I}\) are bounded by \(\eta\). We bound the error from below by \(\lambda_{\min}(A_\perp)^{1/2}\) times the norm of its projection onto the surfaces in \(\mathcal{I}\), which is at least \(N_\theta^{1/2}(|\bar x^*_i|-\eta)_+\) on each surface. \(\square\)

The proof of Theorem 1 gives \(\|x^*-u_0\|_A\le\|r_0\|_{L^+}\alpha^{-1/2}\), and \(\|u_0\|_A=(u_0^TA_\perp u_0)^{1/2}\) does not depend on \(\chi_\parallel\) because \(Lu_0=0\). Therefore \(\|x^*\|_A\) is bounded uniformly in \(\chi_\parallel\), \(\eta=O(\alpha^{-1/2})\), and \(x^*\) converges to the flux function \(u_0\). If \(u_0\) is nonzero on at least one surface in \(\mathcal{I}\), the right-hand side of (13) tends to a positive constant as \(\chi_\parallel\to\infty\). Equation (12) shows that, at finite \(\chi_\parallel\), the surface mean of every vector in the support on a surface with a missing node is at most of order \(\alpha^{-1/2}\). Missing a single node of a flux surface therefore suppresses the flux-function component on that surface, and in the limit the surface is lost. In the computational basis, the support must therefore contain every node of every surface that carries energy. In a non-aligned discretization, the field lines pass near every node, and the support must cover essentially the whole space.

\hypertarget{v.-numerical-results}{%
\section*{V. Numerical results}}

\begin{figure}
\centering
\includegraphics[width=1\textwidth,height=\textheight]{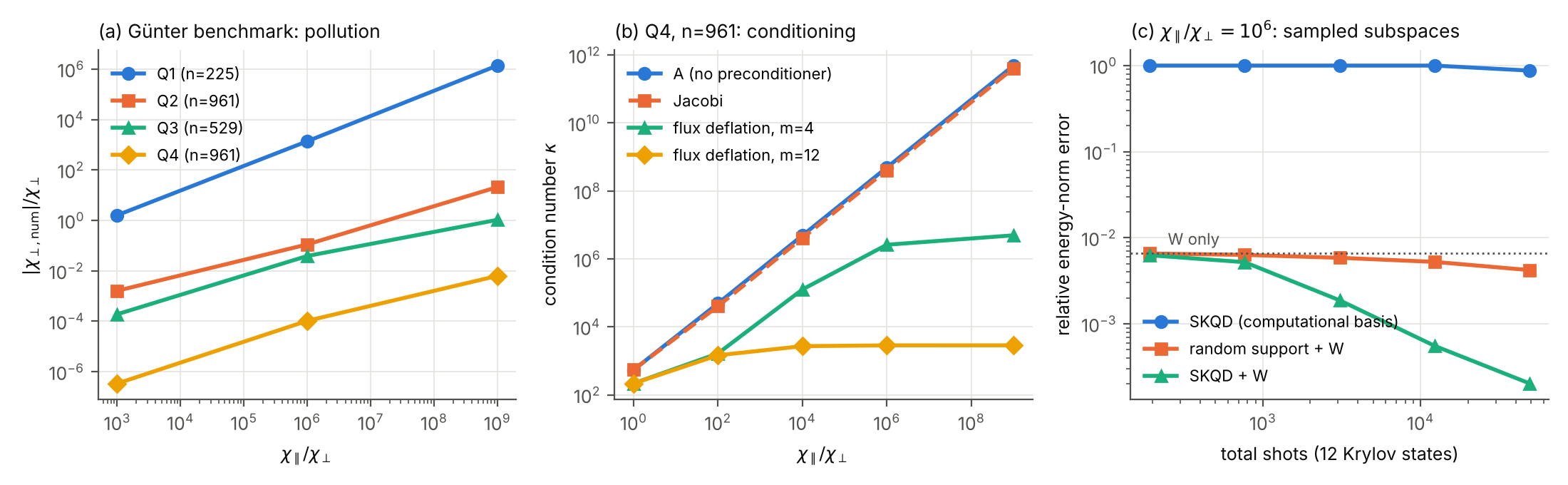}
\caption{Non-aligned discretization of the Günter benchmark. (a) Numerical perpendicular diffusion for elements of degree 1--4 with about \(10^3\) unknowns. (b) Condition number of the degree-4 system with \(n=961\), without preconditioning, with Jacobi preconditioning, and after deflation of a flux-function space of dimension \(m\). (c) Energy-norm error of Krylov sampling at \(\chi_\parallel/\chi_\perp=10^6\) in the computational basis, with a flux-function space \(W\) added, and with a random support of the same size plus \(W\).}
\end{figure}

\hypertarget{a.-non-aligned-discretization}{%
\subsection*{A. Non-aligned discretization}}

Figure 2(a) shows the numerical perpendicular diffusion at \(\chi_\parallel/\chi_\perp=10^9\). With about \(10^3\) unknowns, linear elements give \(\chi_{\perp,\mathrm{num}}/\chi_\perp = 1.4\times10^6\) (\(n=225\)), quadratic elements give \(21\) (\(n=961\)), and degree-4 elements give \(6.3\times10^{-3}\) (\(n=961\)). High-order elements are therefore required at the problem sizes accessible to present quantum hardware.

Figure 2(b) shows that the condition number of the degree-4 system is proportional to \(\chi_\parallel/\chi_\perp\) and reaches \(4.9\times10^{11}\) at \(10^9\). Jacobi preconditioning does not change this behavior. On the other hand, deflation of a space spanned by twelve powers of \(\psi\) reduces the effective condition number to about \(2.9\times10^3\) for every ratio from \(10^4\) to \(10^9\). The number of deflated conjugate-gradient iterations at \(10^9\) is 310, compared with 5352 without deflation.

Figure 2(c) shows the first indication of the barrier. At \(\chi_\parallel/\chi_\perp = 10^6\), Galerkin solutions on supports sampled from twelve Krylov states remain at an error of 0.88 even with \(4.9\times10^4\) shots, when the support covers 84\% of the nodes. Adding the flux-function space reduces the error to \(2.0\times10^{-4}\), which is twenty times smaller than for a random support of the same size. The sampling therefore carries information, but the computational basis alone cannot use it.

\hypertarget{b.-field-aligned-discretization}{%
\subsection*{B. Field-aligned discretization}}

\begin{figure}
\centering
\includegraphics[width=1\textwidth,height=\textheight]{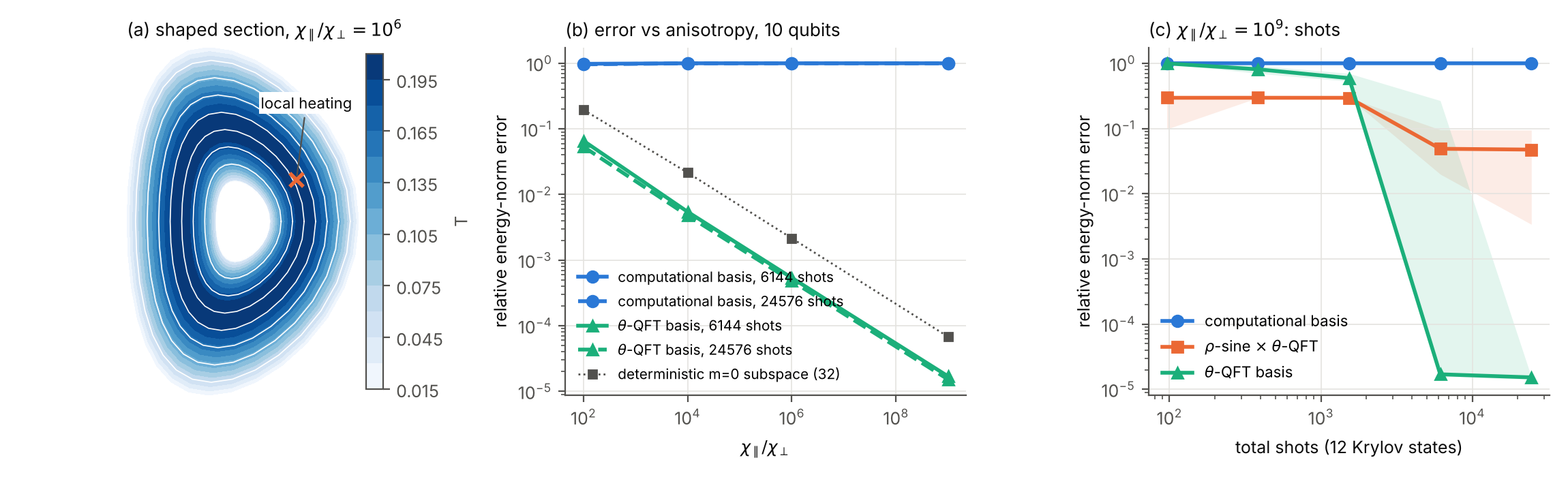}
\caption{Field-aligned discretization of the shaped section (2) with ten qubits. (a) Temperature at \(\chi_\parallel/\chi_\perp=10^6\); the white curves are flux surfaces and the cross marks the localized source. (b) Energy-norm error versus anisotropy for the computational basis and the \(\theta\)-QFT basis at two shot budgets, and for the deterministic \(m=0\) block. (c) Error versus the total number of shots at \(\chi_\parallel/\chi_\perp=10^9\); the bands show the range over five seeds.}
\end{figure}

Figure 3 shows the results for the shaped section (2). The condition number grows from \(6.5\times10^3\) at \(\chi_\parallel/\chi_\perp=10^2\) to \(6.4\times10^{10}\) at \(10^9\). In the computational basis, the error is 0.96--1.0 for every ratio and every shot budget up to \(2.5\times10^4\). In the \(\theta\)-QFT basis with \(6.1\times10^3\) shots, the error is \(5.5\times10^{-3}\), \(5.4\times10^{-4}\), and \(1.7\times10^{-5}\) at \(10^4\), \(10^6\), and \(10^9\). The error decreases as \((\chi_\perp/\chi_\parallel)^{1/2}\), in agreement with Theorem 1.

Figure 3(c) shows that the error drops sharply between \(1.5\times10^3\) and \(6.1\times10^3\) shots. This threshold corresponds to the collection of all 32 states of the \(m=0\) block, and it does not depend on \(\chi_\parallel\). The sampled subspace is more accurate than the deterministic \(m=0\) block by a factor of about four, because the samples also capture part of the \(m\neq0\) correction.

We also tested a sine transform on the \(\rho\) register in addition to the \(\theta\)-QFT. This combination stalls at an error of \(4.7\times10^{-2}\), because the \(m=0\) profile spreads over many sine modes with small probabilities. A Chebyshev grid in \(\rho\) has a related difficulty: the node next to the inner boundary has a small amplitude and is often missed, and its absence causes a perpendicular kink. The uniform \(\rho\) grid with direct measurement of the \(\rho\) register avoids both difficulties.

\begin{figure}
\centering
\includegraphics[width=1\textwidth,height=\textheight]{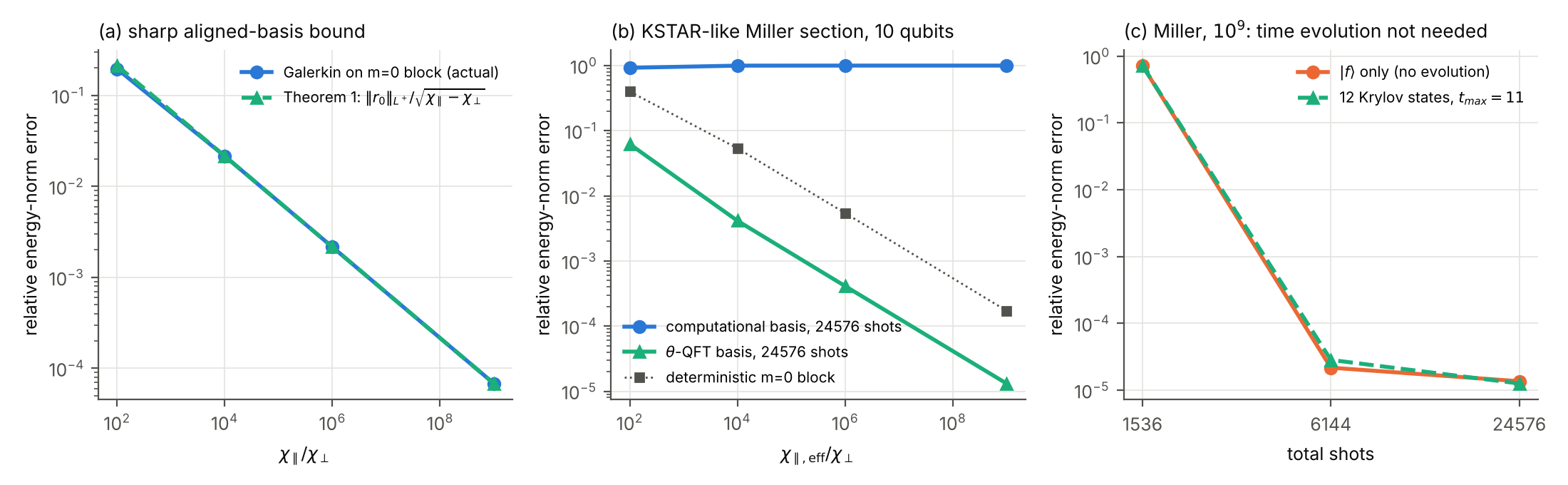}
\caption{(a) Theorem 1: Galerkin error on the \(m=0\) block and the bound (9) for the shaped section. (b) KSTAR-like Miller section (3) with ten qubits and \(2.5\times10^4\) shots. (c) Error versus the number of shots at \(\chi_\parallel/\chi_\perp=10^9\) for sampling \(|f\rangle\) only and for twelve Krylov states.}
\end{figure}

\hypertarget{c.-verification-of-the-theorems}{%
\subsection*{C. Verification of the theorems}}

Table I compares the Galerkin error on the \(m=0\) block with the bound (9). The ratio of the bound to the actual error is 1.088, 1.002, 1.0001, and 1.0000 at \(10^2\), \(10^4\), \(10^6\), and \(10^9\). The ratio approaches one as \(1+O(\alpha^{-1})\), as predicted by Corollary 1. The inequality (12) was verified on the computational-basis Galerkin solutions at all four ratios. For example, at \(10^9\) the largest surface mean of \(x_S\) on an incomplete surface is \(1.0\times10^{-7}\), below the bound \(1.8\times10^{-4}\). At \(\chi_\parallel/\chi_\perp=10^{12}\) the condition number reaches \(6\times10^{13}\), and the double-precision reference solution is no longer reliable. We therefore report results up to \(10^9\).

Figure 5 shows the verification for both sections with the sources of (7). Figures 5(a) and 5(b) show the solutions at \(\chi_\parallel/\chi_\perp=10^9\). In both sections, the temperature is constant on each flux surface within the resolution of the figure, although the source is localized. The peak lies on the surfaces that pass through the heating spot. Figure 5(c) shows the ratio of the bound (9) to the actual error minus one. For the Miller section, \(\chi_\perp\) depends on \(\rho\), and we take \(A_\perp\) at \(\chi_\parallel=\max\chi_\perp=5\), so that \(\alpha=\chi_\parallel-5\). The ratio is 1.27, 1.011, 1.0006, and 1.00001 for the Miller section at \(\alpha=10^2\), \(10^4\), \(10^6\), and \(10^8\). For both sections, the deviation from one decreases by a factor that approaches ten per decade of \(\alpha\), in agreement with Corollary 1. At \(\alpha=10^9\), the deviation is below \(10^{-6}\), which is the precision of the calculation described below, and it is not shown.

\begin{figure}
\centering
\includegraphics[width=1\textwidth,height=\textheight]{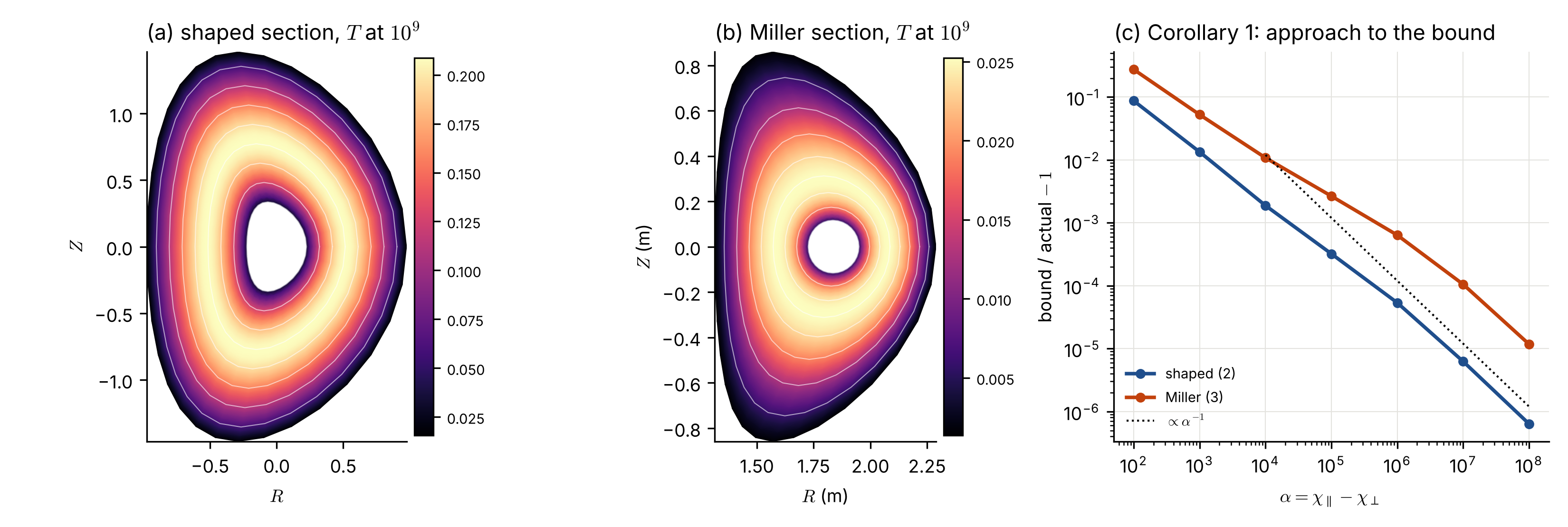}
\caption{Verification for the shaped section (2) and the Miller section (3) with the sources of (7) and \(N_\rho=N_\theta=32\). (a), (b) Temperature at \(\chi_\parallel/\chi_\perp=10^9\); white curves are flux surfaces. (c) Ratio of the bound (9) to the actual error of the \(m=0\) block, minus one, versus \(\alpha\). For the Miller section, \(A_\perp\) is taken at \(\chi_\parallel=5\). The dotted line shows the slope \(\alpha^{-1}\) of Corollary 1.}
\end{figure}

We checked the numerical precision at \(10^9\), where the condition number is \(6.4\times10^{10}\) for the shaped section and \(2.4\times10^{10}\) for the Miller section. All errors are computed from the difference \(x^*-x_S\) in double precision. We repeated the calculation with three steps of iterative refinement, with residuals evaluated in extended precision (64-bit mantissa), for both \(x^*\) and \(x_S\). The refinement changes \(x^*\) by \(6\times10^{-8}\) in the relative energy norm. For the supports used in this paper, the relative change of the reported error is below \(10^{-6}\) for the \(m=0\) block, below \(4\times10^{-5}\) for errors of order \(10^{-5}\), and about 3\% for errors of order \(3\times10^{-7}\). Errors below \(10^{-7}\) are therefore not resolved, and we do not interpret them. The alternative form \(\|x^*-x_S\|_A^2=x^{*T}f-x_S^Tf\) suffers from cancellation at \(10^9\) and is not used.

\begin{longtable}[]{@{}
  >{\raggedright\arraybackslash}p{(\columnwidth - 8\tabcolsep) * \real{0.2000}}
  >{\raggedright\arraybackslash}p{(\columnwidth - 8\tabcolsep) * \real{0.2000}}
  >{\raggedright\arraybackslash}p{(\columnwidth - 8\tabcolsep) * \real{0.2000}}
  >{\raggedright\arraybackslash}p{(\columnwidth - 8\tabcolsep) * \real{0.2000}}
  >{\raggedright\arraybackslash}p{(\columnwidth - 8\tabcolsep) * \real{0.2000}}@{}}
\caption{Galerkin error on the \(m=0\) block and the bound of Theorem 1 for the shaped section.}\tabularnewline
\toprule\noalign{}
\begin{minipage}[b]{\linewidth}\raggedright
\(\chi_\parallel/\chi_\perp\)
\end{minipage} & \begin{minipage}[b]{\linewidth}\raggedright
\(10^2\)
\end{minipage} & \begin{minipage}[b]{\linewidth}\raggedright
\(10^4\)
\end{minipage} & \begin{minipage}[b]{\linewidth}\raggedright
\(10^6\)
\end{minipage} & \begin{minipage}[b]{\linewidth}\raggedright
\(10^9\)
\end{minipage} \\
\midrule\noalign{}
\endfirsthead
\toprule\noalign{}
\begin{minipage}[b]{\linewidth}\raggedright
\(\chi_\parallel/\chi_\perp\)
\end{minipage} & \begin{minipage}[b]{\linewidth}\raggedright
\(10^2\)
\end{minipage} & \begin{minipage}[b]{\linewidth}\raggedright
\(10^4\)
\end{minipage} & \begin{minipage}[b]{\linewidth}\raggedright
\(10^6\)
\end{minipage} & \begin{minipage}[b]{\linewidth}\raggedright
\(10^9\)
\end{minipage} \\
\midrule\noalign{}
\endhead
\bottomrule\noalign{}
\endlastfoot
actual error & \(1.95\times10^{-1}\) & \(2.15\times10^{-2}\) & \(2.15\times10^{-3}\) & \(6.81\times10^{-5}\) \\
bound (9) & \(2.12\times10^{-1}\) & \(2.15\times10^{-2}\) & \(2.15\times10^{-3}\) & \(6.81\times10^{-5}\) \\
bound/actual & 1.088 & 1.002 & 1.0001 & 1.0000 \\
\end{longtable}

\hypertarget{d.-kstar-like-geometry}{%
\subsection*{D. KSTAR-like geometry}}

Figure 4(b) shows the results for the Miller section (3). The condition number reaches \(2.4\times10^{10}\) at \(10^9\). The computational basis gives an error of 0.93--1.0 for every ratio. The \(\theta\)-QFT basis with \(2.5\times10^4\) shots gives \(4.1\times10^{-3}\), \(4.1\times10^{-4}\), and \(1.3\times10^{-5}\) at \(10^4\), \(10^6\), and \(10^9\). The shot threshold lies between \(6.1\times10^3\) and \(2.5\times10^4\). It is higher than for the shaped section because the peaked source and the toroidal factor reduce the probability of the outermost surfaces. It still does not depend on \(\chi_\parallel\).

\hypertarget{e.-time-evolution}{%
\subsection*{E. Time evolution}}

Figure 4(c) compares sampling of the right-hand side state \(|f\rangle\) alone (\(K=1\)) with twelve Krylov states (\(t_{\max}=11\)). The two give the same error within the seed-to-seed spread for both geometries and for the six-qubit instance. In the aligned basis, \(|f\rangle\) already assigns nonzero probability to every state of the \(m=0\) block, because the source has a nonzero average on every flux surface. The Krylov evolution adds no new support.

\hypertarget{vi.-comparison-with-other-sample-based-methods}{%
\section*{VI. Comparison with other sample-based methods}}

We tested the other sample-based methods on the degree-4 Günter benchmark (\(n=961\)). The results are summarized in Table II.

\textbf{SQD.} We sampled from the exact lowest eigenvector, which is the most favorable case. In the computational basis at \(\chi_\parallel/\chi_\perp=10^9\), a subspace of 512 states (53\% of the space) captures 93\% of the probability but overestimates the lowest eigenvalue by a factor of 330. In a two-dimensional Fourier basis, 64 states capture a probability of \(0.999998\), and the eigenvalue is still overestimated by a factor of \(8.8\times10^4\). The reason is that a missing amplitude \(\varepsilon\) costs \(\varepsilon^2\) in probability but \(\chi_\parallel\varepsilon^2\) in energy. A sample-based subspace must therefore resolve a tail probability of order \(\chi_\perp/\chi_\parallel\).

\textbf{Code-space recovery and support expansion.} Code-space recovery repairs noisy samples by detecting code-space violations of a dual-rail encoding {[}11{]}. The barrier above is present for noiseless samples, so that recovery does not remove it. We also applied the reference-guided expansion of Ref. {[}11{]}, which adds single- and double-bit-flip mutants to the support. In the computational basis, the error becomes small only when the expansion covers all 961 states. In the two-dimensional Fourier basis at \(10^9\), the expanded support of 405 states still gives an energy-norm error of 1.0.

\textbf{Residual-norm linear solver.} The sample-based linear solver of Ref. {[}9{]} obtains the solution as the ground state of the projected cost Hamiltonian \(A(\mathsf{I}-|b\rangle\langle b|)A\). Even with the full support, this formulation computed in double precision gives an error of 0.82 at \(10^6\) and 1.0 at \(10^9\), whereas the energy-norm Galerkin solution is exact. In exact arithmetic, the ground state of the cost Hamiltonian is the solution. The cost Hamiltonian, however, squares the condition number to about \(2\times10^{23}\), which exceeds the resolution of double precision, and the failure is a failure in finite precision. The energy-norm formulation of Ref. {[}10{]} projects \(A\) itself and does not have this defect.

\begin{longtable}[]{@{}
  >{\raggedright\arraybackslash}p{(\columnwidth - 4\tabcolsep) * \real{0.3333}}
  >{\raggedright\arraybackslash}p{(\columnwidth - 4\tabcolsep) * \real{0.3333}}
  >{\raggedright\arraybackslash}p{(\columnwidth - 4\tabcolsep) * \real{0.3333}}@{}}
\caption{Sample-based methods on the degree-4 Günter benchmark at \(\chi_\parallel/\chi_\perp=10^9\) (\(n=961\)).}\tabularnewline
\toprule\noalign{}
\begin{minipage}[b]{\linewidth}\raggedright
method and basis
\end{minipage} & \begin{minipage}[b]{\linewidth}\raggedright
support
\end{minipage} & \begin{minipage}[b]{\linewidth}\raggedright
result
\end{minipage} \\
\midrule\noalign{}
\endfirsthead
\toprule\noalign{}
\begin{minipage}[b]{\linewidth}\raggedright
method and basis
\end{minipage} & \begin{minipage}[b]{\linewidth}\raggedright
support
\end{minipage} & \begin{minipage}[b]{\linewidth}\raggedright
result
\end{minipage} \\
\midrule\noalign{}
\endhead
\bottomrule\noalign{}
\endlastfoot
SQD, computational basis, ideal state & 512 states (53\%) & eigenvalue overestimated by 330 \\
SQD, 2D Fourier basis, ideal state & 64 states & eigenvalue overestimated by \(8.8\times10^4\) \\
SQD, 2D Fourier basis, ideal state & 256 states & eigenvalue exact; linear-solve error \(1.8\times10^{-5}\) \\
bit-flip expansion, computational basis, \(10^4\) shots & 958 states & linear-solve error 0.46 \\
bit-flip expansion, 2D Fourier basis, \(10^4\) shots & 405 states & linear-solve error 1.0 \\
residual-norm solver, full support & 961 states & linear-solve error 1.0 (\(10^6\): 0.82) \\
energy-norm Galerkin, full support & 961 states & exact \\
\end{longtable}

\hypertarget{vii.-hardware-realization}{%
\section*{VII. Hardware realization}}

Table III summarizes the resources. The normalized operator \(\hat A\) has about \(10^4\) Pauli terms for ten qubits and about \(4\times10^2\) for six qubits, with a one-norm of 3.1 and 2.0. A qDRIFT simulation {[}14{]} of \(e^{-i\hat At}\) for \(t=11\) with error 0.1 requires about \(2.9\times10^5\) and \(6.5\times10^4\) CNOT gates. These depths exceed present hardware. Since the aligned measurement requires no time evolution (Sec. V E), the circuit reduces to the preparation of \(|f\rangle\) followed by the QFT on the \(\theta\) register. Generic amplitude encoding requires about \(2^n\) CNOT gates {[}15{]}, and the QFT on five qubits requires about 26. The total is about 75 CNOT gates for six qubits and about \(1.05\times10^3\) for ten qubits.

\begin{longtable}[]{@{}
  >{\raggedright\arraybackslash}p{(\columnwidth - 4\tabcolsep) * \real{0.3333}}
  >{\raggedright\arraybackslash}p{(\columnwidth - 4\tabcolsep) * \real{0.3333}}
  >{\raggedright\arraybackslash}p{(\columnwidth - 4\tabcolsep) * \real{0.3333}}@{}}
\caption{Resource estimates.}\tabularnewline
\toprule\noalign{}
\begin{minipage}[b]{\linewidth}\raggedright
item
\end{minipage} & \begin{minipage}[b]{\linewidth}\raggedright
6 qubits (\(8\times8\))
\end{minipage} & \begin{minipage}[b]{\linewidth}\raggedright
10 qubits (\(32\times32\))
\end{minipage} \\
\midrule\noalign{}
\endfirsthead
\toprule\noalign{}
\begin{minipage}[b]{\linewidth}\raggedright
item
\end{minipage} & \begin{minipage}[b]{\linewidth}\raggedright
6 qubits (\(8\times8\))
\end{minipage} & \begin{minipage}[b]{\linewidth}\raggedright
10 qubits (\(32\times32\))
\end{minipage} \\
\midrule\noalign{}
\endhead
\bottomrule\noalign{}
\endlastfoot
Pauli terms (99.9\% of the one-norm) & 351--419 & 10109--11772 \\
one-norm of \(\hat A\) & 2.0 & 3.1 \\
qDRIFT CNOT count, \(t=11\) & \(6.5\times10^4\) & \(2.9\times10^5\) \\
QFT on \(\theta\) register & about 9 & about 26 \\
preparation of \(\lvert f\rangle\) & about 64 & about 1024 \\
proposed circuit (no evolution) & about 75 & about \(1.05\times10^3\) \\
separable \(\lvert f\rangle\) (Sec. VII), logical CNOT & --- & about 80 \\
transpiled on ibm\_cleveland (CZ gates) & 91 (computational), 101 (\(\theta\)-QFT) & 75 (computational), 109 (\(\theta\)-QFT), separable \(\lvert f\rangle\) \\
\end{longtable}

\hypertarget{a.-experiment}{%
\subsection*{A. Experiment}}

The main experiment uses ten qubits, the same size as the numerical results of Sec. V B. The section is (2), with \(N_\rho=N_\theta=32\) and the operators of Sec. II C for \(\chi_\parallel/\chi_\perp=10^2\), \(10^4\), \(10^6\), and \(10^9\). The generic preparation of a ten-qubit state requires about \(10^3\) CNOT gates (Table III). We therefore use a separable load vector,
\[
f_{ij}=g(\rho_i)\,h(\theta_j),\qquad g(\rho)=1+2e^{-(\rho-0.6)^2/(2\cdot0.08^2)},\qquad h(\theta)=1+3e^{-(\theta-0.8)^2/(2\cdot0.4^2)}.
\]
This load represents a heating source localized near \(\rho=0.6\) and \(\theta=0.8\) on a uniform background. Since \(|f\rangle=|g\rangle\otimes|h\rangle\), the state is prepared by two five-qubit blocks, one on each register. The logical circuit then requires about 80 CNOT gates. The two registers are not entangled. This does not affect the test, because Theorems 1 and 2 concern the measurement basis and not the preparation of \(|f\rangle\).

The \(\theta\) register consists of qubits 0--4 and the \(\rho\) register of qubits 5--9. Two circuits are prepared. The first prepares \(|f\rangle\) and measures in the computational basis. The second prepares \(|f\rangle\), applies the QFT to the \(\theta\) register, and measures. The final swaps of the QFT are omitted, and the \(\theta\) index of each outcome is bit-reversed in post-processing. Since \(f\) does not depend on \(\chi_\parallel\), the same two circuits serve all anisotropy ratios. Each measured state \((i,k)\) of the second circuit is mapped to the two real Fourier vectors with \(|m|=\min(k,N_\theta-k)\) on the flux surface \(\rho_i\).

The circuits were run on ibm\_cleveland on October 22, 2026. Before submission, the logical circuits were checked against the exact outcome distributions to \(10^{-12}\). The circuits were transpiled with optimization level 3, and the layout with the fewest two-qubit gates out of 20 seeds was kept. The transpiled circuits contain 75 and 109 CZ gates (Table III). Gate twirling and dynamical decoupling were enabled. No readout correction and no post-selection were applied. A pilot run of 2048 shots for each circuit was followed by a main run of six separate jobs of 8192 shots, alternating between the two circuits. The billed QPU times were 3 s for the pilot run and 27 s for the main run.

For a shot budget \(B\), we draw \(B\) outcomes without replacement from the pooled record of \(2.46\times10^4\) shots, solve the energy-norm Galerkin problem on the observed support, and repeat 200 times. We report the median and the quartiles of the relative energy-norm error. The quartiles describe the variability of subsamples drawn from one pooled record. They are not confidence intervals, and the 200 subsamples are not independent hardware runs. The independent units are the jobs, and their reproducibility is examined separately below. As a control, we replace the observed support by the same number of basis vectors chosen uniformly in the same basis. This control separates the effect of the measurement basis from the effect of the support size. The same analysis is applied to each job separately in order to check the reproducibility. The linear algebra in the subspace is done classically.

\hypertarget{b.-ten-qubit-results}{%
\subsection*{B. Ten-qubit results}}

\begin{figure}
\centering
\includegraphics[width=1\textwidth,height=\textheight]{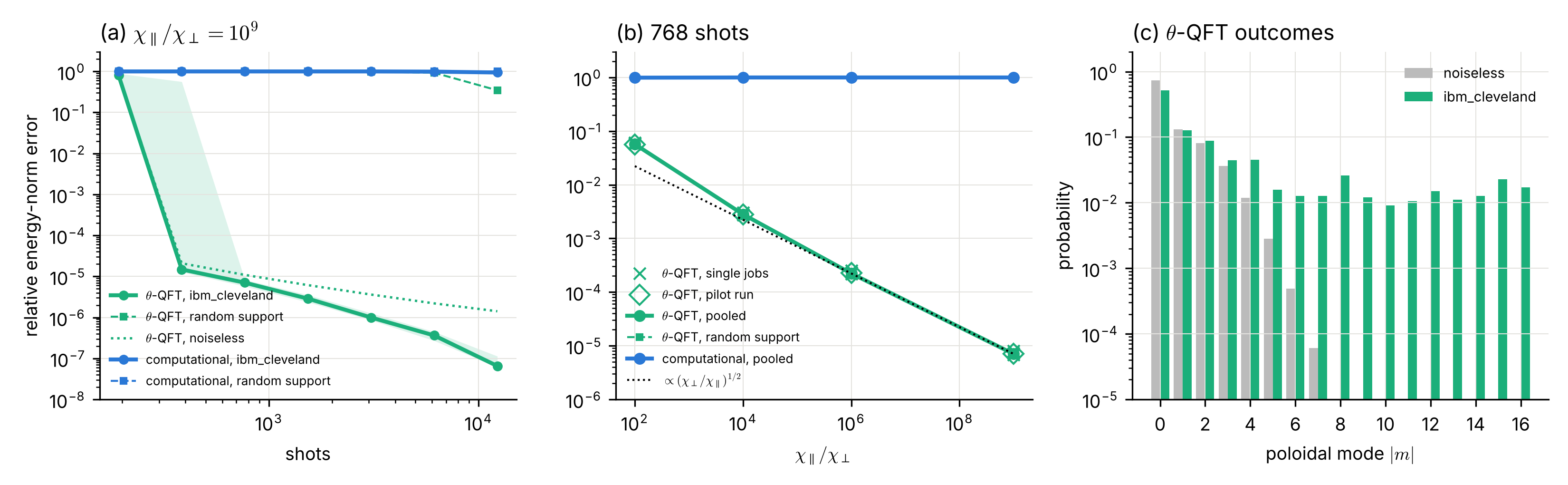}
\caption{Ten-qubit experiment on ibm\_cleveland. (a) Energy-norm error versus the shot budget at \(\chi_\parallel/\chi_\perp=10^9\). The bands show the quartiles over 200 subsamples of the pooled record of three jobs; they describe the subsampling variability, not independent repetitions. Dashed lines use a random support of the same size, and the dotted line uses noiseless samples. The random-support control in the computational basis coincides with the computational-basis curve. (b) Error at 768 shots versus anisotropy for the pooled record, for each of the three \(\theta\)-QFT jobs, and for the pilot run. The random-support control for the \(\theta\)-QFT basis lies under the computational-basis curve. (c) Probability of each poloidal mode \(|m|\) in the \(\theta\)-QFT outcomes.}
\end{figure}

Figure 6(a) shows the error at \(\chi_\parallel/\chi_\perp=10^9\). In the \(\theta\)-QFT basis, the error is \(1.5\times10^{-5}\) with 384 shots, \(7.0\times10^{-6}\) with 768 shots, and \(3.6\times10^{-7}\) with 6144 shots (with a numerical uncertainty of a few percent, Sec. V C). A random support of the same size gives 0.999, 0.998, and 0.93. In the computational basis, the error is 1.00 up to 3072 shots, 0.99 with 6144 shots, and 0.94 with \(1.2\times10^4\) shots, although 993 of the 1024 states are observed at the last budget. This is the barrier of Theorem 2 at the size used in Sec. V. Missing nodes on a few flux surfaces are sufficient to lose those surfaces, and the number of missing nodes decreases only slowly with the number of shots.

Figure 6(b) shows the error at 768 shots as a function of the anisotropy. The \(\theta\)-QFT error is \(5.7\times10^{-2}\), \(2.8\times10^{-3}\), \(2.2\times10^{-4}\), and \(7.0\times10^{-6}\) for \(\chi_\parallel/\chi_\perp=10^2\), \(10^4\), \(10^6\), and \(10^9\). The ratios between successive points are 12.6 from \(10^4\) to \(10^6\) and 31.7 from \(10^6\) to \(10^9\), which are consistent with the \((\chi_\perp/\chi_\parallel)^{1/2}\) scaling of Theorem 1, for which the ratios are 10 and 31.6. At \(10^2\), the error is not yet in the asymptotic regime. The computational-basis error is between 0.99 and 1.00 at all four ratios. At \(10^9\), the three \(\theta\)-QFT jobs give \(7.9\times10^{-6}\), \(7.3\times10^{-6}\), and \(6.8\times10^{-6}\), and the pilot run gives \(7.1\times10^{-6}\). The computational-basis jobs give 0.98--1.00 with 6144 shots each. The result is therefore reproducible from job to job.

Figure 6(c) shows how the device noise enters. The noiseless distribution puts 73.6\% of the probability on \(m=0\). On the device, this fraction is 51.7\%, and the remainder is spread over all \(|m|\). A depolarizing fit gives an effective circuit fidelity of 0.57 for the \(\theta\)-QFT circuit and 0.74 for the computational-basis circuit. However, all 32 states of the \(m=0\) block are observed in every \(\theta\)-QFT job. The median number of shots needed to observe all of them is 340. Once the \(m=0\) block is in the support, Theorem 1 applies, and the additional vectors contributed by the noise can only reduce the error. This is why the hardware curve in Fig. 6(a) lies below the noiseless curve. The effect comes from the larger support and is not an advantage of the noise. A random support of the same size still gives an error of order unity.

\hypertarget{c.-six-qubit-results}{%
\subsection*{C. Six-qubit results}}

\begin{figure}
\centering
\includegraphics[width=1\textwidth,height=\textheight]{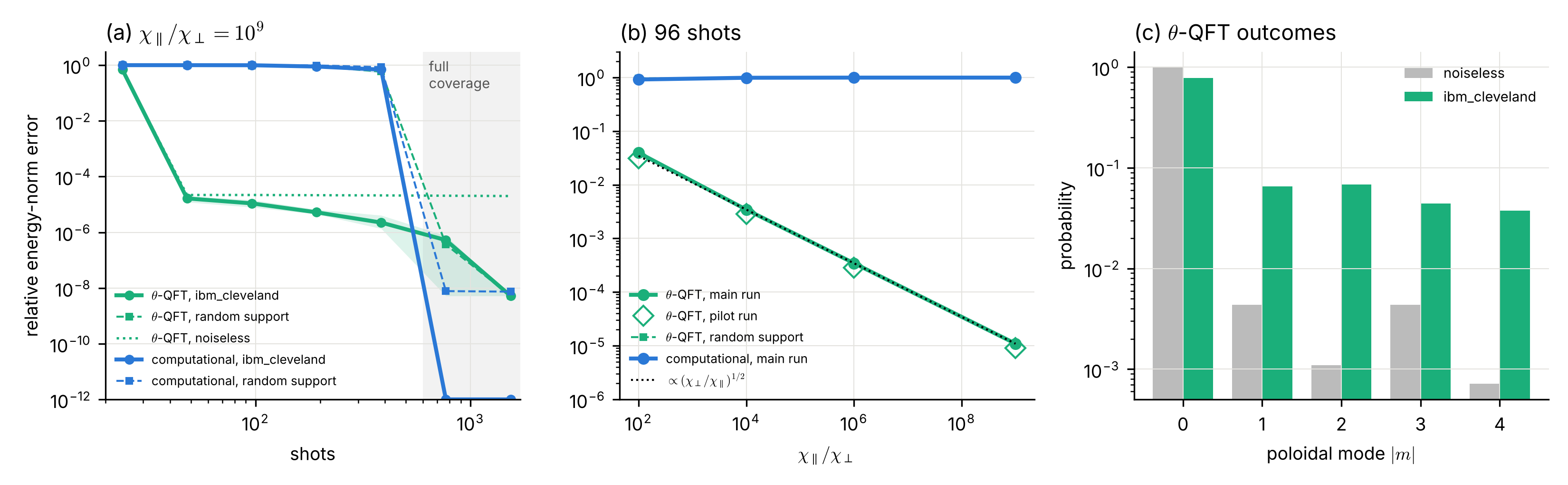}
\caption{Six-qubit experiment on ibm\_cleveland with the shaped section and the localized source of Sec. V B. (a) Energy-norm error versus the shot budget at \(\chi_\parallel/\chi_\perp=10^9\). In the shaded region, the support covers nearly all 64 states. (b) Error at 96 shots versus anisotropy. (c) Probability of each poloidal mode \(|m|\) in the \(\theta\)-QFT outcomes.}
\end{figure}

We also ran a six-qubit instance with \(N_\rho=N_\theta=8\) and the non-separable source of Sec. V B. The generic preparation of \(|f\rangle\) is used, and the transpiled circuits contain 91 and 101 CZ gates. The main run consists of three repetitions of 8192 shots for each circuit, and the billed QPU time was 16 s. Figure 7 summarizes the results. With 96 shots at \(\chi_\parallel/\chi_\perp=10^9\), the \(\theta\)-QFT basis gives \(1.1\times10^{-5}\), a random support gives 0.96, and the computational basis gives 0.99. The \(\theta\)-QFT error at 96 shots is consistent with the \((\chi_\perp/\chi_\parallel)^{1/2}\) scaling from \(10^4\) to \(10^9\). The effective circuit fidelity is 0.69. In this instance, the computational-basis support covers all 64 states above about 700 shots, and both errors then drop to the round-off level. The ten-qubit experiment removes this limitation.

\hypertarget{d.-trapped-ion-processor-and-the-kstar-like-geometry}{%
\subsection*{D. Trapped-ion processor and the KSTAR-like geometry}}

The ten-qubit circuits of Sec. VII A were also run on the trapped-ion processor IonQ Forte-1 through the IonQ v0.4 API, with debiasing disabled. The logical circuits are the same as on ibm\_cleveland. They were submitted in terms of \(R_x\), \(R_y\), \(R_z\), and CNOT gates and compiled by the provider to the native gate set. Since the device has all-to-all connectivity, no routing is needed. According to the job metadata, the provider-compiled circuits contain 52 and 72 two-qubit gates. The API returns probabilities rather than individual shots. In our data, every probability of a scientific job is a multiple of the inverse number of shots, and we reconstruct the counts as the probability times the number of shots. The order of the shots within a job is not available, which does not affect the analysis, since the subsampling depends only on the multiset of outcomes. The outcomes are returned as bitstrings that start with qubit 0 and omit trailing zeros. As in Ref. {[}12{]}, we determined the bit order with diagnostic circuits. Three diagnostic circuits that flip qubit 0, qubit 9, and qubits 0 and 5 give the outcomes 1, 512, and 33 in our convention, with probabilities of 1.00, 0.98, and 0.98. The remaining probability of 0.02 in two diagnostic jobs is not reported by the API and is left unassigned. A pilot run of 1024 shots for each circuit was followed by a main run of six separate jobs of 2048 shots, alternating between the two circuits. The main run was carried out on October 22--23, 2026, and the execution time was 122 s for each computational-basis job and 146 s for each \(\theta\)-QFT job. A depolarizing fit to the pooled main-run record gives an effective circuit fidelity of 0.84 for the \(\theta\)-QFT circuit and 0.93 for the computational-basis circuit.

Since \(f\) does not depend on the geometry, the circuits of Sec. VII A also serve the Miller section of Sec. V D. The geometry enters only through the operator \(A\) in the classical Galerkin step. We therefore analyze each hardware record with both operators. The results for the two operators are therefore not independent hardware experiments. This analysis is possible because the \(m=0\) block is the kernel of \(L\) for any geometry with nested flux surfaces, so that the aligned measurement does not depend on the shape of the surfaces.

\begin{figure}
\centering
\includegraphics[width=0.85\textwidth,height=\textheight]{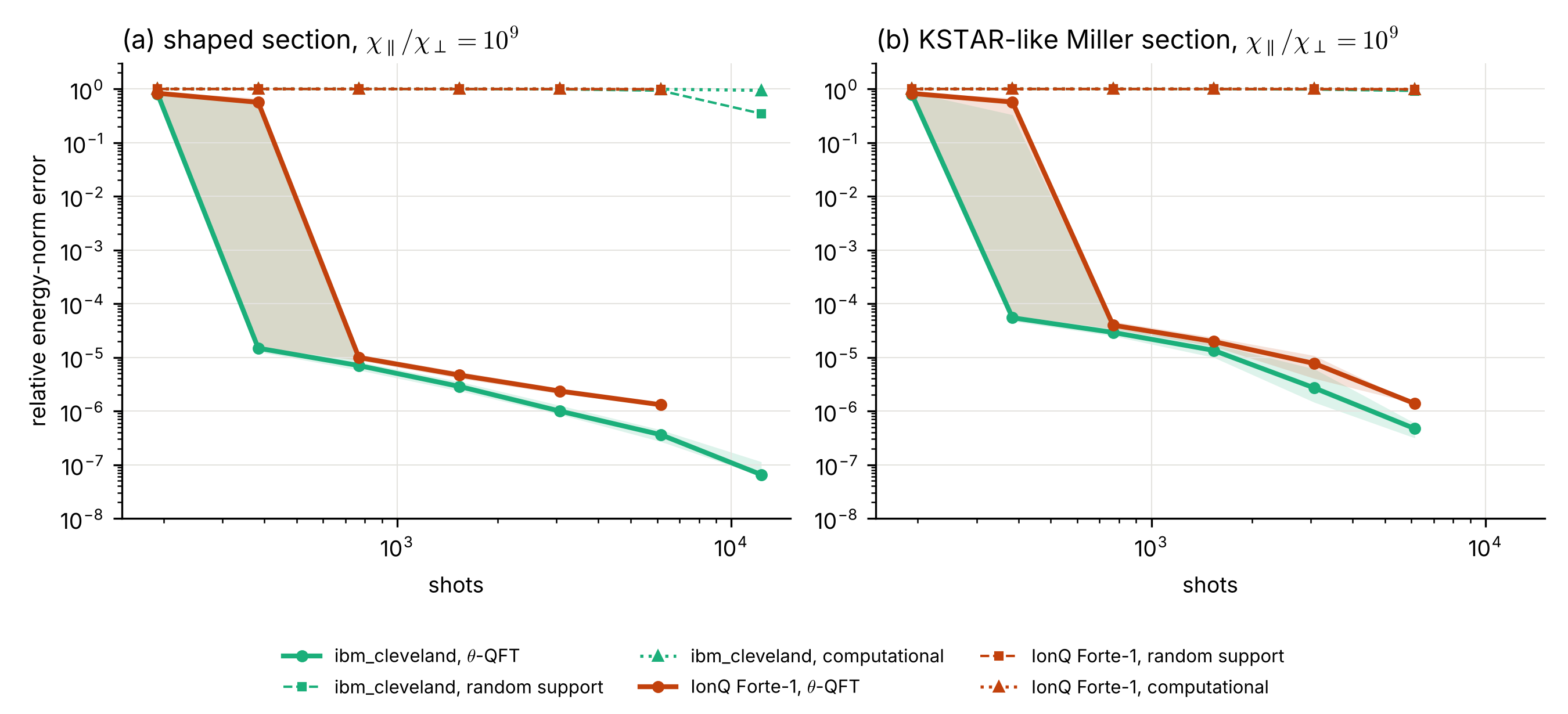}
\caption{Ten-qubit experiments on ibm\_cleveland (\(2.46\times10^4\) shots for each circuit) and IonQ Forte-1 (\(6.1\times10^3\) shots for each circuit) at \(\chi_\parallel/\chi_\perp=10^9\). (a) Shaped section (2). (b) KSTAR-like Miller section (3), evaluated on the same records. Solid lines: \(\theta\)-QFT basis, with quartile bands that describe the subsampling variability of the pooled record. Dashed lines: random support of the same size in the \(\theta\)-QFT basis. Dotted lines: computational basis.}
\end{figure}

\begin{longtable}[]{@{}
  >{\raggedright\arraybackslash}p{(\columnwidth - 8\tabcolsep) * \real{0.2000}}
  >{\raggedright\arraybackslash}p{(\columnwidth - 8\tabcolsep) * \real{0.2000}}
  >{\raggedright\arraybackslash}p{(\columnwidth - 8\tabcolsep) * \real{0.2000}}
  >{\raggedright\arraybackslash}p{(\columnwidth - 8\tabcolsep) * \real{0.2000}}
  >{\raggedright\arraybackslash}p{(\columnwidth - 8\tabcolsep) * \real{0.2000}}@{}}
\caption{Ten-qubit results at \(\chi_\parallel/\chi_\perp=10^9\): median error over subsamples of 768 shots drawn from the pooled main-run record. The pooled record contains three jobs for each circuit on each device. ibm\_cleveland: CZ gates, gate twirling and dynamical decoupling. IonQ Forte-1: provider-native gates, debiasing disabled. The two rows for each device use the same hardware data.}\tabularnewline
\toprule\noalign{}
\begin{minipage}[b]{\linewidth}\raggedright
device
\end{minipage} & \begin{minipage}[b]{\linewidth}\raggedright
operator
\end{minipage} & \begin{minipage}[b]{\linewidth}\raggedright
\(\theta\)-QFT
\end{minipage} & \begin{minipage}[b]{\linewidth}\raggedright
random support
\end{minipage} & \begin{minipage}[b]{\linewidth}\raggedright
computational
\end{minipage} \\
\midrule\noalign{}
\endfirsthead
\toprule\noalign{}
\begin{minipage}[b]{\linewidth}\raggedright
device
\end{minipage} & \begin{minipage}[b]{\linewidth}\raggedright
operator
\end{minipage} & \begin{minipage}[b]{\linewidth}\raggedright
\(\theta\)-QFT
\end{minipage} & \begin{minipage}[b]{\linewidth}\raggedright
random support
\end{minipage} & \begin{minipage}[b]{\linewidth}\raggedright
computational
\end{minipage} \\
\midrule\noalign{}
\endhead
\bottomrule\noalign{}
\endlastfoot
ibm\_cleveland & shaped section (2) & \(7.0\times10^{-6}\) & 0.998 & 1.00 \\
ibm\_cleveland & Miller section (3) & \(2.9\times10^{-5}\) & 0.997 & 1.00 \\
IonQ Forte-1 & shaped section (2) & \(9.9\times10^{-6}\) & 0.999 & 1.00 \\
IonQ Forte-1 & Miller section (3) & \(4.0\times10^{-5}\) & 0.999 & 1.00 \\
\end{longtable}

Figure 8 and Table IV compare the two devices. On both devices and for both operators, the \(\theta\)-QFT basis gives a median error of order \(10^{-5}\) for subsamples of 768 shots, whereas a random support of the same size and the computational basis give errors of order unity. On IonQ, the random-support control stays above 0.99 and the computational basis above 0.98 up to 6144 shots. The three IonQ \(\theta\)-QFT jobs give \(9.9\times10^{-6}\), \(9.8\times10^{-6}\), and \(9.5\times10^{-6}\) for the shaped section and \(3.8\times10^{-5}\) for the Miller section in each job. On IonQ, the \(\theta\)-QFT error at 768 shots decreases by factors of 11.9 and 31.8 from \(10^4\) to \(10^6\) and from \(10^6\) to \(10^9\) for the shaped section, and by 12.2 and 31.5 for the Miller section. These ratios are consistent with the scaling of Theorem 1 over the tested range. With the Miller operator, the error is larger by a factor of about four on both devices. Part of this factor comes from the geometry, since the error of the \(m=0\) block alone is \(1.0\times10^{-4}\) for the Miller section and \(4.2\times10^{-5}\) for the shaped section.

The two devices differ in the number of shots needed to reach the low-error regime. The median number of shots needed to observe all 32 states of the \(m=0\) block is 445 on IonQ and 340 on ibm\_cleveland. With 384 shots, the median error on IonQ is therefore still 0.56, whereas it is \(1.5\times10^{-5}\) on ibm\_cleveland. Although the IonQ distribution is closer to the noiseless one, it is also more concentrated. It contains 543 distinct outcomes in 6144 shots, and the noise contributes fewer additional states. Beyond 768 shots, the IonQ curve follows the noiseless curve, while the ibm\_cleveland curve lies below it because of the additional support contributed by the noise. The pilot run on IonQ gave an effective fidelity of 0.66 and the same error at 768 shots within a factor of 1.2.

\hypertarget{e.-discussion}{%
\subsection*{E. Discussion}}

The hardware results are consistent with both theorems. The aligned basis needs only the \(m=0\) block, which a noisy device still delivers within a few hundred shots. In the computational basis, the noise does not help, because the error is set by the last missing node rather than by the number of observed states. The effective circuit fidelity of the \(\theta\)-QFT circuit is lower than that of the computational-basis circuit, since it contains more gates. The two devices differ in connectivity, native gates, and noise, and the \(\theta\)-QFT circuit has an effective fidelity of 0.57 on one and 0.84 on the other. Even so, the qualitative contrast between the two bases persists on both tested platforms. The comparison shows the robustness of the result across platforms. It is not a comparison of the platforms, since the native gates, compilation, and error suppression differ.

The shot cost of the aligned basis is set by the smallest probability of a state in the \(m=0\) block. For the ten-qubit load, this probability is \(p_{\min}=8.3\times10^{-3}\) for noiseless sampling, and the estimate of Sec. IV A gives \(p_{\min}^{-1}\ln(N_\rho/\epsilon)=500\) shots for \(\epsilon=1/2\). The median number of shots needed to observe the whole block is 380 for noiseless samples, 340 on ibm\_cleveland, and 445 on IonQ Forte-1. The noise therefore changes the shot cost only by a factor of order one, in two opposite ways. It moves probability out of the \(m=0\) block, which increases the cost. It also makes the probabilities within the block more uniform, which decreases the cost. On ibm\_cleveland, the second effect dominates, and on IonQ the first. In both cases, the cost does not depend on \(\chi_\parallel\), because neither the load nor the circuit does.

The contrast with the computational basis is a matter of structure rather than of probability. In the computational basis, the \(m=0\) block is spread over all \(N_\theta\) nodes of each surface, and Theorem 2 requires every node of every surface that carries energy. For the ten-qubit load, the support contains 993 of the 1024 states after \(1.2\times10^4\) shots on ibm\_cleveland, and the error is still 0.94. The missing states are few, but each of them removes a surface. A larger number of shots reduces the number of missing states only slowly, because the last states are the least probable ones. In the aligned basis, one outcome per surface is sufficient, and the number of required outcomes is reduced from \(N_\rho N_\theta\) to \(N_\rho\).

The aligned subspace plays the same role as the coarse space of deflation and asymptotic-preserving methods {[}4{]}. These classical methods remove the near-null space of the parallel operator explicitly, so that the remaining problem is well conditioned. The aligned measurement does the same within a sample-based method, without any change of the quantum circuit except a QFT on the poloidal register. The result is not specific to the shaped section. Since the \(m=0\) block is the kernel of \(L\) for any geometry with nested flux surfaces, the same measured support serves both the shaped and the KSTAR-like section, as Table IV shows.

The error that remains after the \(m=0\) block is collected is of order \((\chi_\perp/\chi_\parallel)^{1/2}\) and is set by the \(m\neq0\) components. Further accuracy requires the \(m\neq0\) modes that carry the response to the localized source. The noisy support already contains some of them, which is why the hardware errors decrease below the bound with the number of shots. A systematic way to reduce this error is iterative refinement, in which the residual is computed classically and a new support is sampled for the correction {[}12{]}. In this approach, the hardware noise sets the rate of convergence rather than the attainable accuracy. Combined with the aligned basis, each correction would start from a subspace that already contains the near-null space.

The present experiments use a separable load, so that the state preparation is shallow. For a general load at ten qubits, the generic preparation requires about \(10^3\) CNOT gates. Loads that are smooth on each flux surface have a small number of significant poloidal modes and can be prepared with fewer gates, for example by a low-rank decomposition between the two registers. The main open problem is three-dimensional fields with islands or stochastic regions, where the near-null space is not known in advance. There, the aligned basis must be constructed from the field-line structure, for example by field-line tracing.

\hypertarget{viii.-limitations}{%
\section*{VIII. Limitations}}

The aligned subspace of Theorem 1 is known before any measurement, and the \(\theta\)-Fourier coefficients of \(f\) can be computed classically in \(O(n\log n)\) operations. For two-dimensional axisymmetric problems with nested flux surfaces, the quantum sampler therefore does not provide an advantage over a classical method that exploits the same structure, such as an asymptotic-preserving scheme {[}4{]}. The results of this paper concern the design of the measurement basis and the limits of sample-based methods, not a computational advantage.

The analysis assumes nested closed flux surfaces and a field tangent to them. Magnetic islands, stochastic regions, and open field lines change the kernel of \(L\). The numerical tests use second-order differences in the aligned case and a single source shape for each geometry. The ten-qubit experiment uses a separable load vector, so the two registers are not entangled. A non-separable load at ten qubits would require about \(10^3\) CNOT gates for the state preparation, which exceeds the present circuit fidelity. The hardware runs on each device were made within one day, and the reproducibility across calibrations has not been tested systematically. Structured state preparation or code-space recovery {[}11{]} may extend the experiment to non-separable loads and larger registers.

\hypertarget{ix.-conclusions}{%
\section*{IX. Conclusions}}

We have studied sample-based quantum subspace solvers for strongly anisotropic heat conduction in magnetized plasmas. In the computational basis, a subspace that misses one node of a flux surface cannot represent the flux-function component on that surface as \(\chi_\parallel\to\infty\), and the error has a floor that does not decrease with the number of shots. The same barrier limits sample-based diagonalization with support expansion, and the residual-norm formulation fails in double precision because it squares the condition number. In a measurement basis aligned with the flux surfaces, the error is bounded by \(\|r_0\|_{L^+}(\chi_\parallel-\chi_\perp)^{-1/2}\), the bound is sharp, and the number of shots does not depend on the anisotropy. We have verified these results on a shaped section and on a KSTAR-like equilibrium with ten qubits, for \(\chi_\parallel/\chi_\perp\) up to \(10^9\). The aligned measurement requires no time evolution, and ten-qubit experiments on a superconducting and a trapped-ion processor are consistent with both theorems at effective circuit fidelities of 0.57 and 0.84.

The general rule is that the measurement basis of a sample-based method must be aligned with the near-null space of the stiff part of the operator. The application of this rule to three-dimensional fields, where the near-null space is not known in advance, remains for future work.

\hypertarget{appendix-a-hardware-jobs}{%
\section*{Appendix A: Hardware jobs}}

\begin{longtable}[]{@{}
  >{\raggedright\arraybackslash}p{(\columnwidth - 6\tabcolsep) * \real{0.2500}}
  >{\raggedright\arraybackslash}p{(\columnwidth - 6\tabcolsep) * \real{0.2500}}
  >{\raggedright\arraybackslash}p{(\columnwidth - 6\tabcolsep) * \real{0.2500}}
  >{\raggedright\arraybackslash}p{(\columnwidth - 6\tabcolsep) * \real{0.2500}}@{}}
\caption{Hardware jobs used in Sec. VII. All jobs were run on October 22--23, 2026. The six-qubit IBM jobs and the ten-qubit IBM pilot job contain both circuits; the other jobs contain one circuit each. The job identifiers are available from the author on request.}\tabularnewline
\toprule\noalign{}
\begin{minipage}[b]{\linewidth}\raggedright
device
\end{minipage} & \begin{minipage}[b]{\linewidth}\raggedright
run
\end{minipage} & \begin{minipage}[b]{\linewidth}\raggedright
circuit
\end{minipage} & \begin{minipage}[b]{\linewidth}\raggedright
shots
\end{minipage} \\
\midrule\noalign{}
\endfirsthead
\toprule\noalign{}
\begin{minipage}[b]{\linewidth}\raggedright
device
\end{minipage} & \begin{minipage}[b]{\linewidth}\raggedright
run
\end{minipage} & \begin{minipage}[b]{\linewidth}\raggedright
circuit
\end{minipage} & \begin{minipage}[b]{\linewidth}\raggedright
shots
\end{minipage} \\
\midrule\noalign{}
\endhead
\bottomrule\noalign{}
\endlastfoot
ibm\_cleveland & 6 qubits, pilot & both bases & 1024 × 2 \\
ibm\_cleveland & 6 qubits, main & both bases, 3 repetitions & 8192 × 6 \\
ibm\_cleveland & 10 qubits, pilot & both bases & 2048 × 2 \\
ibm\_cleveland & 10 qubits, main & computational, rep. 1 & 8192 \\
ibm\_cleveland & 10 qubits, main & \(\theta\)-QFT, rep. 1 & 8192 \\
ibm\_cleveland & 10 qubits, main & computational, rep. 2 & 8192 \\
ibm\_cleveland & 10 qubits, main & \(\theta\)-QFT, rep. 2 & 8192 \\
ibm\_cleveland & 10 qubits, main & computational, rep. 3 & 8192 \\
ibm\_cleveland & 10 qubits, main & \(\theta\)-QFT, rep. 3 & 8192 \\
IonQ Forte-1 & bit-order diagnostic & X on qubit 0 & 100 \\
IonQ Forte-1 & bit-order diagnostic & X on qubit 9 & 100 \\
IonQ Forte-1 & bit-order diagnostic & X on qubits 0 and 5 & 100 \\
IonQ Forte-1 & 10 qubits, pilot & computational & 1024 \\
IonQ Forte-1 & 10 qubits, pilot & \(\theta\)-QFT & 1024 \\
IonQ Forte-1 & 10 qubits, main & computational, rep. 1 & 2048 \\
IonQ Forte-1 & 10 qubits, main & \(\theta\)-QFT, rep. 1 & 2048 \\
IonQ Forte-1 & 10 qubits, main & computational, rep. 2 & 2048 \\
IonQ Forte-1 & 10 qubits, main & \(\theta\)-QFT, rep. 2 & 2048 \\
IonQ Forte-1 & 10 qubits, main & computational, rep. 3 & 2048 \\
IonQ Forte-1 & 10 qubits, main & \(\theta\)-QFT, rep. 3 & 2048 \\
\end{longtable}

\hypertarget{acknowledgments}{%
\section*{Acknowledgments}}

This work was supported by Korea National Research Foundation (NRF) grant No.~NRF-2026-25612745, ICT R\&D program of IITP-2023-2021-0-01810, RS-2023-00225385, AFOSR grant FA2386-22-1-4052. The work was partly supported by the Cleveland Clinic Quantum Start-up Catalyzer Program and Amazon Web Services. The author acknowledges the use of IBM Quantum services for this work through the Cleveland Clinic's Quantum Start-up Catalyzer Program. The views expressed are those of the author, and do not reflect the official policy or position of IBM or the IBM Quantum team. This work was also supported by the National Quantum Laboratory at Maryland (QLab). The author acknowledges QLab for providing quantum computing resources on IonQ devices for this work. The views expressed are those of the author, and do not reflect the official policy or position of IonQ. The author is a co-founder of Singularity Quantum Inc.~and holds an equity interest in the company. The author used the AI assistants ChatGPT (OpenAI) and Claude (Anthropic) to help draft and edit the manuscript, to check the mathematical derivations, and to write numerical and analysis scripts. The author reviewed and verified all results and takes full responsibility for the content of this article.

\hypertarget{author-declarations}{%
\section*{Author declarations}}

\textbf{Conflict of interest.} The author is a co-founder of Singularity Quantum Inc.~and holds an equity interest in the company.

\hypertarget{data-availability}{%
\section*{Data availability}}

The data that support the findings of this study are available from the corresponding author upon reasonable request. These include the per-shot records of the IBM Quantum jobs, the outcome probabilities returned by the IonQ jobs, the problem instances, and the analysis code.

\hypertarget{references}{%
\section*{References}}

\begin{enumerate}
\def\labelenumi{\arabic{enumi}.}
\tightlist
\item
  S. I. Braginskii, in \emph{Reviews of Plasma Physics}, edited by M. A. Leontovich (Consultants Bureau, New York, 1965), Vol. 1, p.~205.
\item
  S. Ichimaru, \emph{Statistical Plasma Physics, Vol. I: Basic Principles} (Addison-Wesley, Redwood City, 1992).
\item
  S. Günter, Q. Yu, J. Krüger, and K. Lackner, J. Comput. Phys. \textbf{209}, 354 (2005).
\item
  P. Degond, A. Lozinski, J. Narski, and C. Negulescu, ``An asymptotic-preserving method for highly anisotropic elliptic equations based on a micro-macro decomposition,'' arXiv:1102.0904.
\item
  I. Y. Dodin and E. A. Startsev, Phys. Plasmas \textbf{28}, 092101 (2021).
\item
  E. Esmaeilifar, D. Ahn, and R. S. Myong, Phys. Fluids \textbf{36}, 106110 (2024).
\item
  J. Robledo-Moreno, M. Motta, H. Haas, A. Javadi-Abhari, P. Jurcevic, W. Kirby, S. Martiel, K. Sharma, S. Sharma, T. Shirakawa, \emph{et al.}, Sci. Adv. \textbf{11}, eadu9991 (2025).
\item
  J. Yu, J. Robledo Moreno, J. T. Iosue, L. Bertels, D. Claudino, B. Fuller, P. Groszkowski, T. S. Humble, P. Jurcevic, W. Kirby, \emph{et al.}, ``Quantum-centric algorithm for sample-based Krylov diagonalization,'' arXiv:2501.09702.
\item
  B. Park, K. Jeong, and D. Ahn, ``A sample-based quantum linear solver through Krylov-state sampling,'' arXiv:2610.23138.
\item
  D. Ahn and B. Park, ``Riemannian formulation of sample-based quantum Krylov solvers for elliptic problems: Energy-norm guarantees and duct-flow applications,'' arXiv:2610.15586.
\item
  B. Park, S. Kang, D. Ahn, and K. Jeong, ``Code-space recovery for sample-based quantum diagonalization beyond native symmetry constraints,'' arXiv:2607.10227.
\item
  D. Ahn, B. Park, R. S. Myong, and F. Klein, ``Quantum--classical iterative refinement for the pressure-Poisson equation on NISQ hardware,'' arXiv:2610.24342.
\item
  R. L. Miller, M. S. Chu, J. M. Greene, Y. R. Lin-Liu, and R. E. Waltz, Phys. Plasmas \textbf{5}, 973 (1998).
\item
  E. Campbell, Phys. Rev.~Lett. \textbf{123}, 070503 (2019).
\item
  V. V. Shende, S. S. Bullock, and I. L. Markov, IEEE Trans. Comput.-Aided Des. Integr. Circuits Syst. \textbf{25}, 1000 (2006).
\end{enumerate}

\end{document}